\documentclass[12pt]{article}
\usepackage{axodraw}
\usepackage{graphicx}
\usepackage{enumerate}
\def\lsim{\mathrel{\vcenter{\hbox{$<$}\nointerlineskip\hbox{$\sim$}}}}
\newcommand{\be}{\begin{equation}}
\newcommand{\ee}{\end{equation}}
\newcommand{\ba}{\begin{eqnarray}}
\newcommand{\ea}{\end{eqnarray}}

\def \N{\sf{I} \hspace{-0.1em}\sf{N}}
\def \unity{{\sf 1} \hspace{-0.1em}{\sf I}}
\def\21{$SU(2) \otimes U(1) $}

\def\lsim{\raise0.3ex\hbox{$\;<$\kern-0.75em\raise-1.1ex\hbox{$\sim\;$}}}
\def\gsim{\raise0.3ex\hbox{$\;>$\kern-0.75em\raise-1.1ex\hbox{$\sim\;$}}}

\newcommand{\mx}{\left[\begin{array}}
\newcommand{\finmx}{\end{array}\right]}
\newcommand{\mxp}{\left(\begin{array}}
\newcommand{\finmxp}{\end{array}\right)}
\def\beq{\begin{equation}}
\def\eeq{\end{equation}}
\def\bea{\begin{eqnarray}}
\def\eea{\end{eqnarray}}

\def\mathbf#1{\hbox{\bf #1}}
\def\textrm#1{\hbox{#1}}

\def\lsim{\raise0.3ex\hbox{$\;<$\kern-0.75em\raise-1.1ex\hbox{$\sim\;$}}}
\def\gsim{\raise0.3ex\hbox{$\;>$\kern-0.75em\raise-1.1ex\hbox{$\sim\;$}}}
\newcommand {\ignore}[1]{}

\begin{document}
\vspace*{-1in}
\renewcommand{\thefootnote}{\fnsymbol{footnote}}
\begin{flushright}
\texttt{
}
\end{flushright}
\vskip 5pt
\begin{center}
{\Large{\bf Constraints on SUSY Seesaw Parameters from
Leptogenesis and Lepton Flavor Violation}} \vskip 25pt

{\sf F. Deppisch$^{1,2}$\footnote[1]{E-mail:
deppisch@physik.uni-wuerzburg.de}, H.
P\"as$^{1,3}$\footnote[2]{E-mail: paes@phys.hawaii.edu}, A.
Redelbach$^4$\footnote[3]{E-mail: A.Redelbach@gsi.de}, R.
R\"uckl$^1$\footnote[4]{E-mail: rueckl@physik.uni-wuerzburg.de}}
\vskip 10pt
{\it \small $^1$Institut f\"ur Theoretische Physik und Astrophysik\\
Universit\"at W\"urzburg\\ D-97074 W\"urzburg, Germany}\\
{\it \small $^2$ Deutsches Elektronen-Synchrotron DESY\\
D-22603 Hamburg, Germany
}\\
{\it \small $^3$~Department of Physics \& Astronomy,\\University
of Hawaii at Manoa,
2505 Correa Road, Honolulu, HI 96822, USA}\\
{\it \small $^4$Gesellschaft f\"ur Schwerionenforschung\\
Planckstra\ss{}e 1, D-64291 Darmstadt, Germany}\\

\vskip 20pt

{\bf Abstract}
\end{center}

\begin{quotation}
{\small We study the constraints on a minimal supersymmetric
seesaw model imposed by neutrino data, charged lepton flavor
violation, thermal leptogenesis and perturbativity. We show that
it is possible to constrain the three heavy Majorana neutrino
masses as well as the complex Yukawa coupling matrix. Our results provide
a first step towards
a seesaw benchmark model for further phenomenological
studies and model building.}
\end{quotation}

\vskip 20pt

\setcounter{footnote}{0}
\renewcommand{\thefootnote}{\arabic{footnote}}

\newpage

\section{Introduction}
The discovery of neutrino masses and mixings in neutrino
oscillation experiments calls for an extension of the Standard
Model (SM) of particle physics. The most elegant way to implement
the required neutrino masses into the SM is via the seesaw
mechanism \cite{seesaw}. Three right-handed neutrinos are added to
the SM particle content, which, being singlets under the SM gauge
symmetry, acquire heavy Majorana masses. By integrating out these
right-handed Majorana neutrinos, light masses for the left-handed
neutrinos are generated naturally. In a supersymmetric version of
this mechanism, the lepton flavor violation present in the
neutrino sector is transmitted also to the sleptons inducing
charged lepton flavor violation (LFV). We thus focus in this work
on a supersymmetric seesaw model. In the framework of degenerate
right-handed neutrinos, LFV processes in the SUSY seesaw model
were studied in \cite{Deppisch:2002vz,Deppisch:2003wt} (see also
\cite{lfv,Hisano:1996cp,Casas:2001sr}
for further phenomenological studies on LFV in
 processes at low energies).

Another particularly attractive feature of the seesaw mechanism is
that it offers a natural possibility to generate the observed
baryon to photon ratio in the universe
\cite{Spergel:2003cb,Tegmark:2003ud},
\begin{equation}\label{eqn:BaryonAsymmetry}
\eta_B=\frac{n_B - n_{\bar{B}}}{n_\gamma}=(6.3\pm 0.3)\cdot 10^{-10}.
\label{etab}
\end{equation}
This process is known as leptogenesis \cite{Yanagida} and utilizes
the fact that Majorana neutrinos do not carry a conserved lepton
number. First, a lepton number asymmetry is generated by
out-of-equilibrium decays of the heavy right-handed Majorana
neutrinos and sneutrinos (in the following subsumed as neutrinos)
in the early universe. This lepton number asymmetry is later on
transformed into a baryon number asymmetry via non-perturbative
Standard Model processes induced by finite temperature transitions
between topologically distinct vacua separated by the sphaleron barrier.
If the right-handed neutrinos
are generated by scattering in the thermal bath, leptogenesis
provides a particularly straightforward prediction of the baryon
asymmetry of the universe. The decay rates of the right-handed
neutrinos and washout processes depend on their masses and Yukawa
couplings. Hence the observed value of $\eta_B$ (\ref{etab}) may
imply constraints on the seesaw model
\cite{Davidson:2002qv,Ellis:2002xg,Pascoli:2003uh}.

In total, the seesaw mechanism introduces 18 new parameters in
addition to the Minimal Supersymmetric Standard Model (MSSM)
parameters. A major drawback of the scheme is, that only 9 of
these parameters describe the three masses, mixing angles and CP
phases of the light neutrinos, which are directly accessible to
the experiment. The other nine parameters are related to the
masses and mixing of the heavy Majorana sector, which may be
related to physics at the GUT scale. The parameters enter the
renormalization group evolution of the neutrino Yukawa couplings
and affect the mixing of the slepton sector, which gives rise to
observable LFV processes such as $l_j\to l_i\gamma$, with
branching ratios that depend on the parameters of the heavy
Majorana sector.

On the other hand, these parameters enter the prediction of the
baryon asymmetry via leptogenesis and thus also affect the lower
bound on the reheating temperature of the universe after
inflation, which governs the gravitino abundance. Since the decay
of long-living gravitinos can dissociate light elements and thus
jeopardize the successful predictions of Big-Bang nucleosynthesis,
a low reheating temperature is desirable.

In this work we study  the constraints on the parameters of the
SUSY seesaw model in a mSUGRA framework imposed by the combined
neutrino data, successful leptogenesis with low reheating
temperature and charged lepton flavor violation. Furthermore, we
require that the Yukawa couplings do not become too large so that
the perturbative theory holds, and assume hierarchical spectra for
both the left- and right-handed neutrinos.

This work is organized as follows. The theoretical framework of
the SUSY seesaw model is outlined in Section 2. Section 3
introduces the neutrino data used in our calculations.
Leptogenesis is explained in Section 4, followed by a discussion
of the gravitino problem in Section 5. Constraints from
leptogenesis and charged lepton flavor violation on the seesaw
parameters are derived in Section 6 and 7.

\section{SUSY seesaw model \label{frame}}
The simplest supersymmetric version of the seesaw mechanism is
described by the superpotential \cite{Hisano:1996cp}
\begin{equation}
\mathcal{W}=\mathcal{W}_{MSSM}+\frac{1}{2}\nu_R^{c\mbox{ }T}M \nu_{R}^c+
            \nu_R^{c\mbox{ }T}Y_{\nu} \; l \cdot h_2,
\end{equation}
where \(\mathcal{W}_{MSSM}\) describes the MSSM, \(\nu_R\) denotes
the right-handed neutrino singlets and \(l\) the left-handed
lepton doublets, $h_{2}$ the Higgs doublet with hypercharge
$+\frac{1}{2}$, and \(Y_\nu\) the matrix of neutrino Yukawa
couplings. Below the mass scale of the lightest right-handed
Majorana neutrino, the effective superpotential
\cite{Casas:2001sr} is obtained by integrating out the heavy
neutrino fields,
\begin{equation}
    \mathcal{W}_{eff} =
        \mathcal{W}_{MSSM} + \frac{1}{2}(Y_\nu \; l\cdot
        h_2)^T M^{-1}(Y_\nu \; l\cdot h_2).
\label{suppot}
\end{equation}
After electroweak symmetry breaking, \(\mathcal{W}_{eff}\)
(\ref{suppot}) leads to the following mass term for the light
neutrinos,
\begin{equation}
    m_{\nu} = m_D^{\mbox{ }T}M^{-1}m_D
        = Y_{\nu}^T M^{-1} Y_{\nu}\langle h^0_2\rangle^2\equiv
        \kappa \langle h^0_2\rangle^2,
\label{massterm}
\end{equation}
where \(\langle h^0_2\rangle^2 = v_2^2\equiv v^2 \sin^2 \beta\) is
the vacuum expectation value (v.e.v.) of the relevant Higgs field
and \(m_D=Y_{\nu}v_2\) is the neutrino Dirac mass matrix. Here it
has been assumed that the scale of \(m_D\) is much smaller than
the scale of the heavy neutrino mass matrix, \(m_D\ll M\). This
corresponds to the so called type I seesaw mechanism, where any
direct (type II) contributions to the light neutrino mass matrix
are neglected. Such terms can be generated, e.g. from v.e.v's of
SU(2)-triplet Higgs fields.

It is convenient to work in the flavor basis in which the charged
lepton Yukawa matrix is diagonal, so that the symmetric matrix
\(\kappa\) is diagonalized by the unitary Maki-Nakagawa-Sakata
(MNS) matrix \(U\),
\begin{equation}
    U^T \kappa U =
        \textrm{Diag}(\kappa_1,\kappa_2,\kappa_3)\equiv D_{\kappa},
\label{mns}
\end{equation}
the light neutrino mass eigenstates being given by
\beq
    m_i=v_2^2\kappa_i, \qquad i=1,2,3.
\eeq If one chooses \(\kappa_i\geq 0\), \(U\) can be written as
\beq\label{eqn:CPphases} U=V\cdot \textrm{Diag}(e^{-i\phi/2},
e^{-i\phi'/2},1), \eeq where \(\phi,\phi'\) are Majorana phases
and \(V\) can be parametrized in the standard
Cabibbo-Kobayashi-Maskawa form, \beq
    V = \left( \begin{array}{ccc}
        c_{13}c_{12} & c_{13}s_{12} & s_{13}e^{-i\delta} \\
        -c_{23}s_{12}-s_{23}s_{13}c_{12}e^{i\delta} & c_{23}c_{12}-
s_{23}s_{13}s_{12}e^{i\delta} & s_{23}c_{13} \\
        s_{23}s_{12}-c_{23}s_{13}c_{12}e^{i\delta} &-s_{23}c_{12}-
c_{23}s_{13}s_{12}e^{i\delta} & c_{23}c_{13}
        \end{array} \right)\label{V_{CKM}}
\eeq
with $c_{ij}=\cos \theta_{ij}$ and $s_{ij}=\sin \theta_{ij}$.

From the relations (\ref{massterm}) and (\ref{mns}), and working
in the basis where \(M\) is diagonal,
\(M=\textrm{Diag}(M_1,M_2,M_3)=D_M\), one obtains
\beq U^*
D_\kappa U^\dagger = (R^T \sqrt{D^{-1}_M} Y_\nu)^T (R^T
\sqrt{D^{-1}_M} Y_\nu),
\eeq
where \(R\) is an arbitrary complex
orthogonal matrix \cite{Casas:2001sr}, \(R^TR=RR^T=1\). The
neutrino Yukawa coupling matrix can thus be expressed as
\beq
    Y_\nu=\sqrt{D_M} R \sqrt{D_{\kappa}} U^{\dagger}.
\label{yuk}
\eeq
The complex orthogonal matrix \(R\) can be
parametrized by three complex angles $\theta_i=x_i + i y_i$:
\beq\label{eqn:rmatrixpara}
    R = \left(\begin{array}{ccc}\label{R_mat_explicit}
    \hat{c}_2 \hat{c}_3 & -\hat{c}_1 \hat{s}_3-\hat{s}_1 \hat{s}_2 \hat{c}_3 &
\hat{s}_1 \hat{s}_3 -\hat{c}_1 \hat{s}_2 \hat{c}_3\\
    \hat{c}_2 \hat{s}_3 & \hat{c}_1 \hat{c}_3-\hat{s}_1 \hat{s}_2 \hat{s}_3  & -
\hat{s}_1 \hat{c}_3 -\hat{c}_1 \hat{s}_2 \hat{s}_3 \\
    \hat{s}_2     & \hat{s}_1 \hat{c}_2              & \hat{c}_1 \hat{c}_2
    \end{array}\right),
\eeq
with \(\hat{c}_i\equiv \cos(\theta_i)\) and \(\hat{s}_i\equiv \sin(\theta_i)\).

We focus on the simplest scenario, where the right-handed
neutrinos are hierarchical\footnote{For a discussion of leptogenesis in scenarios with degenerate
neutrinos see \cite{Pilaftsis:1998pd,Hambye:2003rt}.},
\be
M_1 \ll M_2 \ll M_3,
\ee
and thus the baryon asymmetry is generated
essentially in the decays of the lightest right-handed neutrino with mass
$M_1$.
Consequently, we also assume a hierarchical left-handed neutrino spectrum,
\be
m_1 \ll m_2 \ll m_3,
\ee
since Yukawa couplings and hierarchical Majorana masses would have to conspire
in an unnatural way to lead to a degenerate spectrum for the light neutrinos.

To zero order in the small ratios $M_1/M_3$, $M_2/M_3$ and $m_1/m_3$
the Yukawa coupling (\ref{yuk}) is given by
\beq
Y_\nu \simeq \frac{\sqrt{M_3 m_3}}{v \sin \beta}
{\rm Diag(0,0,1)}\cdot R \cdot {\rm Diag} \left(0,\sqrt{\frac{m_2}{m_3}}e^{i
\phi'/2},1\right)
V^\dagger,
\label{yzord}
\eeq
with $m_2\simeq\sqrt{\Delta m^2_{12}}$ and $m_3\simeq\sqrt{\Delta m^2_{23}}$,
defined
through the mass squared differences measured in neutrino oscillation
experiments.
Inserting the parametrization (\ref{R_mat_explicit}) for $R$ one finally arrives
at
\bea
(Y_\nu)_{ij} &\simeq& \frac{\sqrt{M_3 m_3}}{v \sin \beta}\hat{c}_2\delta_{i3}
\left(\hat{s}_1 \sqrt{\frac{m_2}{m_3}} e^{i\phi'/2} V_{j2}^* +\hat{c}_1
V_{j3}^* \right).
\label{yprox}
\eea

Before proceeding, we briefly summarize
the renormalization group evolution of neutrino Yukawa
couplings from \(m_Z\) to the GUT-scale \(M_{GUT}\) for
non-degenerate seesaw scales \cite{Antusch:2002rr}.
Below \(M_1\), the heavy Majorana neutrinos are decoupled, so that
\(Y_{\nu}=0\) and only the effective light neutrino mass matrix
\(\kappa\) evolves, starting from the input value
\(\kappa(m_Z)=U^* D_{\kappa} U^\dagger\). At the
\(M_1\)-threshold, the corresponding right-handed Majorana
neutrino becomes active and one has
\beq
    (Y_\nu)_{ij}\Big{|}_{M_1+}= \delta_{i1}\left(
    \sqrt{D_M} R \sqrt{D_{\kappa}} U^{\dagger}\right)_{ij}\Big{|}_{M_1-},
\eeq
where \(M_1\!\!+ (M_1-)\) denotes the right (left) limit
approaching the scale \(M_1\). Furthermore, using
(\ref{massterm}), the tree-level matching condition for \(\kappa\)
at \(M_1\) is given by \beq
    (\kappa)_{ij}\Big{|}_{M_1+}= (\kappa)_{ij}\Big{|}_{M_1-}
    -\left(Y_\nu^{\mbox{ }T}\right)_{i1}
    \frac{1}{M_1}\left(Y_\nu\right)_{1j}\Big{|}_{M_1-}.
\eeq

Note that the masses of heavy neutrinos also run above the respective
mass thresholds.

Now we continue with the evolution from \(M_1\) to \(M_2\) using
the input values \(\kappa|_{M_1+}\), \(Y_\nu|_{M_1+}\) and
\(M_1|_{M_1+}\). The matching
\ba
    (Y_\nu)_{ij}\Big{|}_{M_2+}&=&
        \left(Y_\nu\right)_{ij}\Big{|}_{M_2-}
        + \delta_{i2}\left(\sqrt{D_M} R \sqrt{D_{\kappa}}
U^{\dagger}\right)_{ij}
        \Big{|}_{M_2-}, \\
    (\kappa)_{ij}\Big{|}_{M_2+}&=& (\kappa)_{ij}\Big{|}_{M_2-}
        - \left(Y_\nu^{\mbox{ }T}\right)_{i2}
        \frac{1}{M_2}\left(Y_\nu\right)_{2j}\Big{|}_{M_2-},
\ea
at \(M_2\) is analogous to the matching at \(M_1\).
Above the \(M_2\) threshold, the \(2\times 2\) submatrix of
\(M\) receives small off-diagonal elements, so that at the \(M_3\)
scale \(M\) has to be diagonalized by an unitary matrix
\(U_M\), \(U_M^T \left.M\right|_{M_3-} U_M = D_M\). The re-diagonalization
of $M$ leads to the redefinition \(Y_\nu\rightarrow U_M^* Y_\nu\).
As has been noted in \cite{Antusch:2002rr}, the renormalization
group equations (RGE's) are invariant under the transformations
that diagonalize $M$.
Therefore the matching conditions at $M_3$ are given by
\ba
    (Y_\nu)_{ij}\Big{|}_{M_3+}&=& \left(U_M^*
        Y_\nu\right)_{ij}\Big{|}_{M_3-}
        + \delta_{i3}\left(\sqrt{D_M} R \sqrt{D_{\kappa}}
        U^{\dagger}\right)_{ij}\Big{|}_{M_3-}, \label{match20}\\
    (\kappa)_{ij}\Big{|}_{M_3+}&=&0. \label{match21}
\ea
Finally, from (\ref{match20}) and (\ref{match21}), we continue the evolution up
to the
unification scale $M_{X}$.

With \(M\) and \(Y_\nu\) now calculated at \(M_X\), we now turn to the slepton
sector,
assuming the mSUGRA universality conditions
\begin{equation}
m^{2}_{\tilde{l}_{L}}=m_{0}^{2}~\unity,\qquad
m^{2}_{\tilde{l}_{R}}=m_{0}^{2}~\unity, \qquad A_e
=A_{0}Y_{l},\qquad A_\nu
=A_{0}Y_{\nu},\label{eqn:mSUGRA_Universality}
\end{equation}
at $M_{X}$, where $m_{0}$ is the common scalar mass and $A_{0}$ the common
trilinear coupling.
At lower scales of order of the SUSY threshold, the mass squared matrix of the
charged sleptons has the form
\begin{equation}\label{ch_slepton_mass_mat}
  m_{\tilde l}^2=\left(
    \begin{array}{cc}
        m^2_{\tilde{l}_{L}}    & m^{2\;\dagger}_{\tilde{l}_{LR}} \\
        m^{2}_{\tilde{l}_{LR}} & m^{2}_{\tilde{l}_{R}}
    \end{array}
      \right),
\end{equation}
where  \(m^2_{\tilde{l}_{L}}\), \(m^{2}_{\tilde{l}_{R}}\) and
\(m^{2}_{\tilde{l}_{LR}}\)
are \(3\times3\) matrices in flavor space,
\(m^2_{\tilde{l}_{L}}\) and \(m^{2}_{\tilde{l}_{R}}\) being hermitian.
The respective matrix elements are given by
\begin{eqnarray}
  (m^2_{\tilde{l}_L})_{ij}     &=& (m_{L}^2)_{ij} + \delta_{ij}\left(m_{l_i}^2 +
m_Z^2 \cos2\beta\left(-\frac{1}{2}+\sin^2\theta_W \right)\right),
\label{slepcorr_1}\\
  (m^2_{\tilde{l}_{R}})_{ij}     &=& (m_{R}^2)_{ij} + \delta_{ij}(m_{l_i}^2 -
m_Z^2 \cos2\beta\sin^2\theta_W), \label{slepcorr_2}\\
 (m^{2}_{\tilde{l}_{LR}})_{ij} &=& (A_e)_{ij}v\cos\beta-
\delta_{ij}m_{l_i}\mu\tan\beta,
\label{slepcorr_3}
\end{eqnarray}
\(\theta_W\) being the weak-mixing
angle, and \(\mu\) the SUSY Higgs-mixing parameter.
The first terms on the r.h.s. of (\ref{slepcorr_1}) - (\ref{slepcorr_3}) receive
the following
contributions:
\begin{eqnarray}
m_{L}^2&=&m_0^2~\unity + (\delta m_{L}^2)_{\textrm{\tiny MSSM}} + \delta
m_{L}^2,
\label{left_handed_SSB} \\
m_{R}^2&=&m_0^2~\unity + (\delta m_{R}^2)_{\textrm{\tiny MSSM}} + \delta
m_{R}^2,
\label{right_handed_SSB}\\
A_e&=&A_0 Y_l+(\delta A_e)_{\textrm{\tiny MSSM}}+\delta A_e \label{A_SSB}.
\end{eqnarray}
Here, $(\delta m^{2}_{L,R})_{\textrm{\tiny MSSM}}$ and $(\delta
A_e)_{\textrm{\tiny MSSM}}$ denote the usual MSSM renormalization
group corrections \cite{RGE,deBoer:1994dg} which are flavor-diagonal.
In addition, the right-handed neutrinos radiatively induce the
flavor non-diagonal terms $\delta m_{L,R}$ and $\delta A_e$,
resulting from the RGE's
\ba
    16\pi^2\frac{d \delta m_{L}^2}{d\ln\mu} &=&
        m_{L}^2 Y_\nu^\dagger Y_\nu + Y_\nu^\dagger Y_\nu m_{L}^2
        + 2\left(Y_\nu^\dagger m_{\tilde{\nu}}^2 Y_\nu+m_{h_2}^2
        Y_\nu^\dagger Y_\nu+A_\nu^\dagger A_\nu\right), \label{mL_RGE_generic}\\
    16\pi^2\frac{d \delta m_{R}^2}{d\ln\mu} &=&
        0, \label{me_RGE_generic}\\
    16\pi^2\frac{d \delta A_e}{d\ln\mu} &=&
        2 Y_e Y_\nu^\dagger A_\nu + A_e Y_\nu^\dagger Y_\nu .
\label{Ae_RGE_generic}
\ea

For qualitative discussions, we mention here the results in the leading
logarithmic approximation,
\begin{eqnarray}\label{eq:rnrges}
  \delta m_{L}^2 &=& -\frac{1}{8 \pi^2}(3m_0^2+A_0^2)(Y_\nu^\dag L Y_\nu),\\
  \delta m_{R}^2 &=& 0,  \\
  \delta A_e &=& -\frac{3 A_0}{16\pi^2}(Y_l Y_\nu^\dag L Y_\nu),
\label{eq:rnrges123}
\end{eqnarray}
with
\beq
L_{ij}=\ln\left(\frac{M_X}{M_{i}}\right)\delta_{ij}.
\eeq

In the following we adopt the particle spectrum of the mSUGRA
benchmark scenario SPS1a \cite{sps1a} as an illustrative example.
This allows us to easily put our work in context with other SUSY
analyses such as \cite{Aguilar-Saavedra:2005pw}. While the
numerical results depend on this choice, the generic features also
apply to other SUSY scenarios with universality conditions such as
in (\ref{eqn:mSUGRA_Universality}).

\section{Neutrino masses and mixing \label{nuparsec}}

An important requirement on the seesaw model is that the observed
light neutrino masses and mixings are reproduced. This is
automatically guaranteed in the $R$ matrix parametrization
sketched in Section~\ref{frame} where the light neutrino
parameters can be used as input at the electroweak scale in the
renormalization group evolution of the Yukawa coupling matrix. In
the following, we use the global best fit values of the neutrino
oscillation parameters obtained in a three neutrino framework
\cite{Maltoni:2003da} (for a more recent analysis, see
\cite{Strumia:2005tc}) comprising KamLAND, CHOOZ, MACRO,
Super-Kamiokande and SNO data as well as the first spectral data
from the K2K long baseline accelerator experiment. The resulting
best fit values are summarized in Tab.~\ref{neutrino_parameters}.
In order to indicate the uncertainties from the experimental
errors one will have to live with, we have used \(2\sigma\)-errors
expected from future measurements as explained in
\cite{Deppisch:2002vz}.

\begin{table}[h]
\centering
\begin{tabular}{|c|c|c|}
\hline
Parameter  & best fit & future range \\
\hline
\hline
\(\sin^2\theta_{23}\)  & 0.52     & \(^{+0.1}_{-0.1}\) \\
\hline
\(\sin^2\theta_{13}\)  & 0.005    & \(^{+0.001}_{-0.005}\) \\
\hline
\(\sin^2\theta_{12}\)  & 0.30     & \(^{+0.05}_{-0.05}\) \\
\hline
\(\Delta m_{12}^2/10^{-5}\textrm{ eV}^2\) & 6.9 & \(^{+0.36}_{-0.36}\) \\
\hline
\(\Delta m_{23}^2/10^{-3}\textrm{ eV}^2\) & 2.3 & \(^{+0.7}_{-0.9}\) \\
\hline
\end{tabular}
\caption{Best-fit values \cite{Maltoni:2003da} and 2$\sigma$ C.L.
uncertainties expected from future measurements of neutrino oscillation
parameters.
\label{neutrino_parameters}}
\end{table}
For the Dirac phase \(\delta\) and the two Majorana phases
\(\phi\) and \(\phi'\) introduced in (\ref{eqn:CPphases}), no
experimental limits exist (compare, however \cite{dps05}).

Upper bounds on the absolute mass scale of neutrinos can be
obtained from tritium beta decay experiments, neutrinoless double
beta decay searches and the neutrino hot dark matter effect on the
cosmological large scale structure and the cosmic microwave
background \cite{Paes:2001nd}. Furthermore, generation of the
baryon asymmetry (\ref{eqn:BaryonAsymmetry}) via thermal
leptogenesis from the decays of the lightest Majorana neutrino
$N_1$ yields a constraint\footnote{An at least mild hierarchy of
right-handed neutrinos is assumed and strong phase cancellations,
which may relax this bound, are neglected  \cite{Hambye:2003rt}.}
on the mass scale of the lightest neutrino
\cite{DiBari:2004en,Buchmuller:new}: \beq m_1<0.15~\mbox{eV}. \eeq
Moreover, as has been mentioned above, a hierarchical right-handed
neutrino spectrum strongly favors a hierarchical left-handed
neutrino spectrum. We therefore assume for consistency that
next-generation double beta decay experiments will not observe
neutrinoless double beta decay and restrict our discussion to
left-handed neutrinos being lighter than the anticipated double
beta decay sensitivity, $m_1 \lsim 0.03$~eV.

\section{Leptogenesis}

Thermal leptogenesis as a mechanism for the generation of the
baryon asymmetry of the universe has been discussed extensively in
\cite{Buchmuller:2004nz,DiBari:2004en}. For the sake of
completeness we review in the following the most important
relations and facts. The baryon to photon ratio generated from
out-of-equilibrium decays of the heavy right-handed neutrinos is
given by
\beq
 \eta_{B}\approx d~a_{\rm Sph}~\epsilon_{1}~\kappa_{f}
    \label{etaBshort}.
\eeq
The $CP$ asymmetry $\epsilon_1$ generated in the decays of the lightest
right-handed neutrino $N_1$
is defined as \cite{Plumacher:1998ex}
\ba
    \epsilon_1&=&\frac{\Gamma\left(N_{1}\to h_2 + l\right)-
          \Gamma (N_{1}\to \bar h_2 + \bar{l})}
          {\Gamma\left(N_{1}\to h_2 + l\right)+
          \Gamma (N_{1}\to \bar h_2 + \bar{l})}.
\ea In the SUSY seesaw model one also has to consider the
supersymmetric versions of these interactions involving e.g. heavy
right-handed sneutrinos \cite{Plumacher:1998ex}. The $CP$
violation in the decays of the \(N_i\) arises from the
interference of tree-level decay diagrams with vertex and
self-energy corrections \cite{Covi:1996wh}
\ba
\epsilon_1 &\simeq&
        -\frac{1}{8\pi}\frac{1}{\left(Y_\nu Y_\nu^\dagger \right)_{11}}
        \sum_{j\neq 1} \Im m\left(\left(Y_\nu Y_\nu^\dagger\right)_{1j}
        \left(Y_\nu Y_\nu^\dagger\right)_{1j}\right)
        f\left(\frac{M_j^2}{M_1^2}\right), \label{epsCP}\\
    f(x)&=&\sqrt{x}\left(\frac{2}{x-1}
        + \ln\frac{1+x}{x}\right).\label{eps_gen}
\ea
For hierarchical heavy Majorana neutrinos \(\frac{M_j^2}{M_1^2}\gg
1\), \(j\neq1\), and apart from strong phase cancellations \cite{Hambye:2003rt},
one obtains \(f\left(\frac{M_j^2}{M_1^2}\right)\simeq
3\frac{M_1}{M_j}\), leading to \cite{Buchmuller:2000as,
Davidson:2002qv}
\ba
    \epsilon_{1} &\simeq&
        -\frac{3}{8\pi}\frac{1}{\left(Y_\nu Y_\nu^\dagger \right)_{11}}
        \sum_{j\neq 1} \Im m\left(\left(Y_\nu Y_\nu^\dagger\right)_{1j}
        \left(Y_\nu Y_\nu^\dagger\right)_{1j}\right)
        \frac{M_1}{M_j}\label{Leptoapproxstandard}.
\label{epsCP2}
\ea
As the MNS matrix \(U\) drops out in (\ref{epsCP2}), it is clear
that non-zero imaginary parts of the
\(R\) matrix elements are necessary to generate a $CP$ asymmetry.

Substituting the expression (\ref{yuk}) into (\ref{epsCP2})
one gets \cite{Davidson:2002qv}
\beq
    \epsilon_1\simeq
        -\frac{3}{8\pi}\frac{M_1}{v_2^2}\frac{\sum_i
        m_i^2 \Im m\left(R_{1i}^2\right)}{\sum_{i} m_i\left|R_{1i}\right|^2}.
\label{epsdi}
\eeq
Using furthermore the orthogonality condition $\sum_{i}R_{1i}^2=1$,
one approximately has \cite{Davidson:2002qv,bound}
\beq
    \left|\epsilon_1\right|\lsim \frac{3}{8\pi}\frac{M_1}{v_2^2}
    \left(m_3-m_1\right).
\label{fourtyfour}
\eeq
For a  hierarchical spectrum of light neutrinos,
the $CP$ asymmetry $\epsilon_1$ is maximal.
On the other hand the relation (\ref{fourtyfour}) also implies a lower bound
on the
$M_1$-scale, e.g. if $\epsilon_1>10^{-6}$, then $M_1> 4\cdot
10^{9}$~GeV \cite{Davidson:2002qv}.

The efficiency factor \(\kappa_{f}\) takes into account the
effects of the washout rate on an initial \((B-L)\) asymmetry by
inverse decays, and by \(\Delta L=1\) and \(\Delta L=2\)
scattering processes. A reliable numerical fit for \(\kappa_{f}\)
for hierarchical light neutrinos in the strong washout regime is
given by
\cite{Buchmuller:2004nz,DiBari:2004en,Hirsch:2001dg,Nielsen:2001fy}
\begin{equation}
    \kappa_{f}(\tilde{m}_{1}) \simeq (1.5 \pm 0.7)\cdot 10^{-2}
\left( \frac{10^{-2}~{\rm eV}}{\tilde{m}_1} \right)^{1.1 \pm
0.1}\!\!\!\!\!\!\!\!\!\!\!\!\!\!. \label{kappafit}
\end{equation}
The generated \((B-L)\) asymmetry
strongly depends on the effective neutrino mass \beq \tilde{m}_{1}
= v^2 \sin^2 \beta \frac{(Y_\nu Y_\nu^\dagger)_{11}}{M_1}. \eeq
For values of $\sqrt{\Delta m^2_{12}}<\tilde{m}_1< \sqrt{\Delta
m^2_{23}}$ in the strong washout regime, the predictions of the
final baryon asymmetry become independent of initial conditions
and have minimal theoretical uncertainties
\cite{Buchmuller:new,DiBari:2004en,DiBari:2002wi}.

Finally one also has to take
into account the dilution of the asymmetry due to standard photon
production from the onset of leptogenesis until the recombination time
of photons \cite{Buchmuller:new}, described by the dilution factor
$d=(3g_\gamma^{rec})/(4g^{MSSM}_\gamma) \simeq 1/78$ in the MSSM,
where \(g_\gamma^{MSSM}\) and \(g^{rec}_\gamma\) denote the degrees of
freedom in the MSSM and at the time of recombination, respectively.

The initial \(B-L\) asymmetry is subsequently converted to a baryon
asymmetry
by sphaleron processes.
In the case of the MSSM with three fermion generations and two Higgs doublets
and taking into account the chemical potentials of all particle species in the
high-temperature phase, one obtains the conversion factor
\beq
    a_{\rm Sph}=\frac{8}{23}\label{Sph_conversion}.
\eeq
Note that this factor of roughly one
third also arises in the SM with one Higgs doublet.

Since $\eta_B$ is known quite accurately and the effective
neutrino mass is given by $\tilde{m}_1=\sum_i
m_i\left|R_{1i}\right|^2$, there is a direct relation between
$M_1$ and the light neutrino masses and $R$ matrix elements, \beq
    M_1 \simeq
        -\frac{8\pi}{3}\frac{\eta_B}{d\,a_{\rm Sph}} \frac{\sum_{i}
m_i\left|R_{1i}\right|^2}{\sum_i m_i^2
        \Im m\left(R_{1i}^2\right)}
        \frac{v_2^2}{\kappa_f}.
    \label{Relation_M1_mi_R}
\eeq
Note that (\ref{Relation_M1_mi_R}) is a direct consequence of a
successful thermal leptogenesis for hierarchical $M_i$.

\section{Gravitino production and reheating temperature
\label{sectrh}}

A major obstacle of thermal leptogenesis can be the generation of
an overabundance of gravitinos which causes serious cosmological
difficulties \cite{GravitinoProblem}. Since the couplings of the
gravitino to ordinary matter are strongly suppressed by the
gravitational scale, it has a very long lifetime. Nevertheless, if
it is heavier than the Lightest Supersymmetric Particle (LSP), it
can decay e.g. radiatively into a photon and a photino. These
decays will occur after the big-bang nucleosynthesis (BBN), unless
the gravitino is heavier than $\sim 10$~TeV
\cite{Hamaguchi:2002vc}. Among the gravitino decay products are
energetic photons which induce electromagnetic cascade processes,
thereby spoiling successful BBN. Since the number of gravitinos
produced during the reheating epoch is approximately proportional
to the reheating temperature $T_R$, one can obtain upper bounds on
$T_R$ depending on the gravitino mass $m_{3/2}$.

According to the analysis \cite{Kawasaki:2000qr}, the upper bounds
corresponding to a heavy, i.e unstable gravitino are given by
$T_R\lsim 10^7,$ $10^9$ and $10^{12}$~GeV for $m_{3/2}=100$~GeV,
1~TeV and 3~TeV, respectively\footnote{More stringent bounds on
the reheating temperature can be found in the recent analysis
\cite{Kohri:2005wn}.}. Assuming that the right-handed neutrinos
are produced thermally after inflation results in a constraint on
$M_1$ \cite{DiBari:2004en}, $M_1<10~T_R$. This potential problem
of thermal leptogenesis can be overcome e.g. in anomaly or gaugino
mediated supersymmetry, as has been realized in
\cite{Gherghetta:1999sw,Buchmuller:2003is}, where the gravitino
either decays before the initiation of nucleosynthesis, or becomes
the stable LSP. Since the gauge couplings decrease above a
critical temperature $T_{*}$, depending on the SUSY breaking
scale, this mechanism leads to a relic gravitino density which is
compatible with the WMAP results and which becomes independent of
the reheating temperature for $T_R>T_{*}$. In any case a small
reheating temperature and Majorana mass $M_1$ are desirable. We
thus study in the following the values of the $R$ matrix in
(\ref{Relation_M1_mi_R}) for which $M_1$ acquires its minimum
value.

\section{Constraints on the $R$ matrix}

As outlined in Section 2, the \(R\) matrix (\ref{eqn:rmatrixpara})
can be parametrized by six real parameters \(x_i\), \(y_i\),
\(i=1,2,3\). With the $x_i$ and $y_i$ given, $M_1$ is fixed by the
leptogenesis condition (\ref{epsdi}). In order not to violate the
gravitino bound on the reheating temperature in mSUGRA models, we require
\(M_1<10^{11}\)~GeV. In addition, $\tilde{m}_1$ has been demanded
to lie in the interval $\sqrt{\Delta
m^2_{12}}<\tilde{m}_1<\sqrt{\Delta m^2_{23}}$, so that the fit
(\ref{kappafit}) for the washout factor $\kappa_{f}$ is
appropriate.

Another important condition for obtaining meaningful results on
physical observables is the reliability of perturbative theory in
the Higgs sector. We therefore require that the largest Yukawa
coupling eigenvalue $\left(Y_{\nu}\right)_{3}$ meets the condition
\(\left|\left(Y_{\nu}\right)_{3}\right|^2/4 \pi \lsim 0.3\). The
largest Yukawa coupling $|\left(Y_{\nu}\right)_3|$ is very
sensitive to the heaviest Majorana mass $M_3$ and the parameters
\(y_i\) in the $R$ matrix. This behavior can be easily understood
from (\ref{yuk}) and the parametrization (\ref{R_mat_explicit})
which imply a $\sqrt{M_3}$ dependence and an exponential
dependence on \(y_i\). Thus the perturbativity bound constrains
$y_i < {\cal O}(1)$. In the subsequent numerical treatment an
upper bound $y_i <1$ has been imposed and remaining points with
values of $\left(Y_\nu\right)_3^2/4\pi>0.3$ have been rejected.

In the right-handed neutrino sector the following assumptions have
been made. The $R$ matrix parameters $x_1$, $x_2$ and $x_3$ are
varied in their full range $[0,2 \pi]$ and the imaginary parts
$y_i$ are varied in the interval $[10^{-3},1]$. For larger values
of \(y_i\), the Higgs sector becomes strongly interacting. On the
other hand, all three parameters $y_i$ have to be non-zero to make
leptogenesis possible.
Finally the hierarchical spectrum of the right-handed neutrinos
has been generated by scattering $M_3$ logarithmically between
$10~M_1$ and $10^4 M_1$ and varying $M_2$ between $M_1$ and
$0.1~M_3$. The condition $M_2 >~ 3 M_1$ required for (42) to hold is met
for the predominant majority of scatter points.

The constraints on the $R$ matrix imposed by the requirement of a
minimal Majorana mass $M_1$ for a given baryon asymmetry can be
read off from (\ref{epsdi}). It is obvious that $M_1$ obtains its
minimum values for $\Im m(R_{1i}) \simeq \Re e(R_{1i})$, and thus
a small $\Re e(R_{1i})$, since the imaginary parts are determined
by the parameters $y_i$. While the contribution of $R_{11}$ in the
sum in (\ref{Relation_M1_mi_R}) is suppressed by the small value
of $m_1$ in hierarchical neutrino spectra, $\Re e(R_{12})$ and
$\Re e(R_{13})$ become small for $x_{2,3} \simeq n \pi$ with $n
\in \N$ independently from the value of \(x_1\), as is obvious
from (\ref{R_mat_explicit}). Thus, the remaining angle \(x_1\) cannot
be constrained in this way. This is physically understandable,
as \(x_1\) is related to the mixing between the right-handed
neutrinos \(\nu_{R_2}\) and \(\nu_{R_3}\), which plays no major
role in the decays of \(\nu_{R_1}\).

\begin{figure}[t]
\centering
\includegraphics[clip,width=0.32\textwidth]{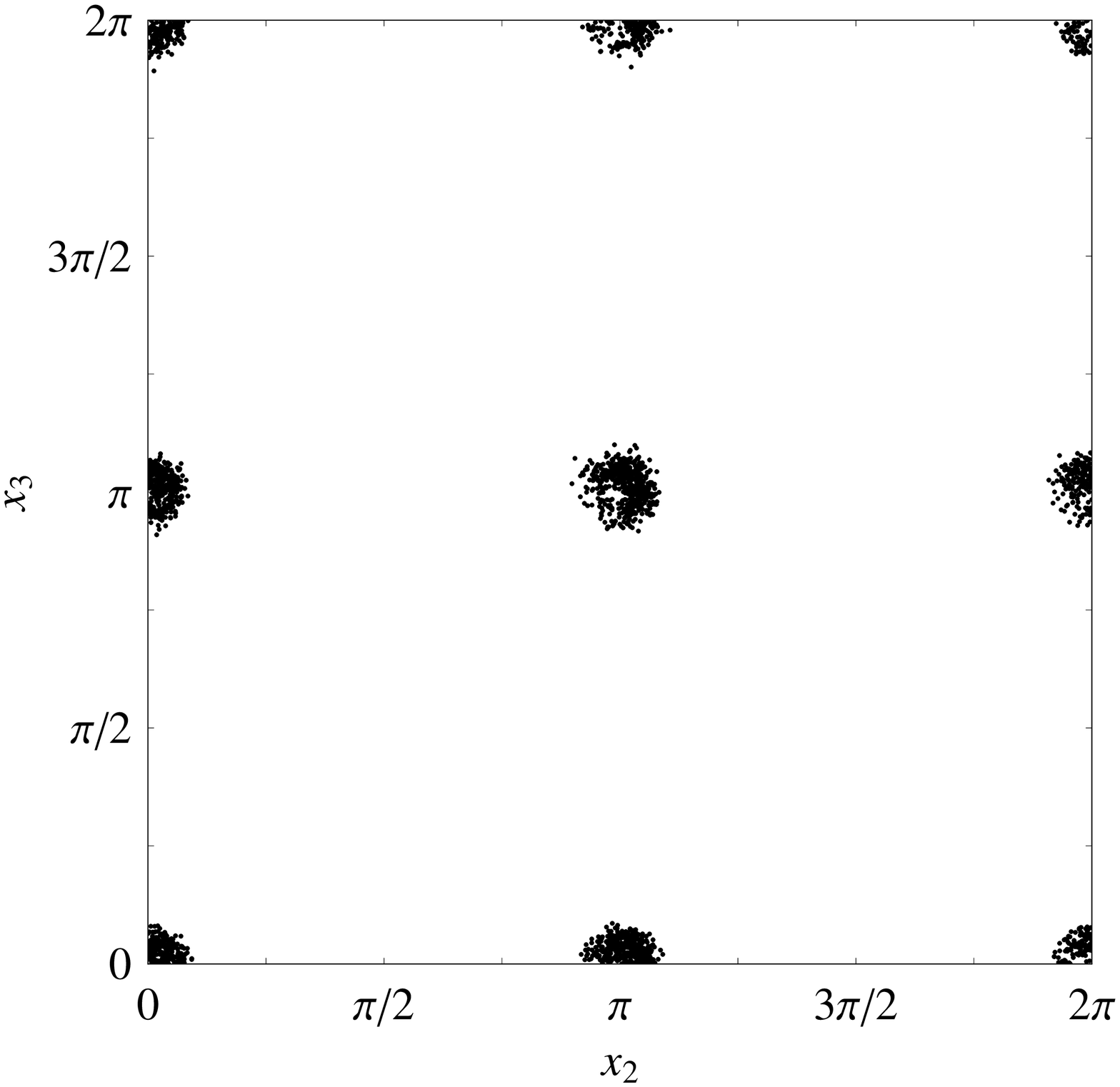}
\includegraphics[clip,width=0.32\textwidth]{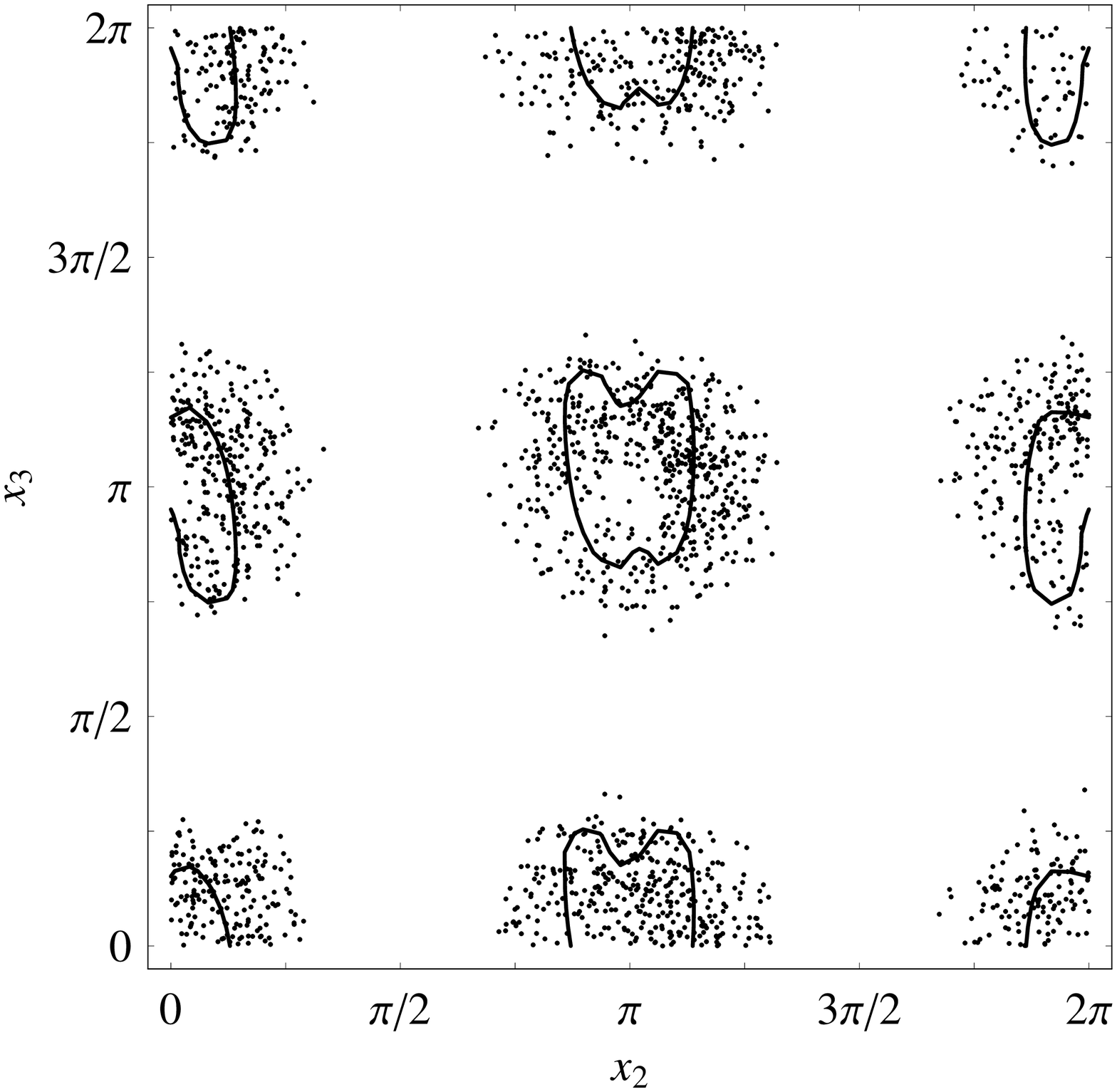}
\includegraphics[clip,width=0.32\textwidth]{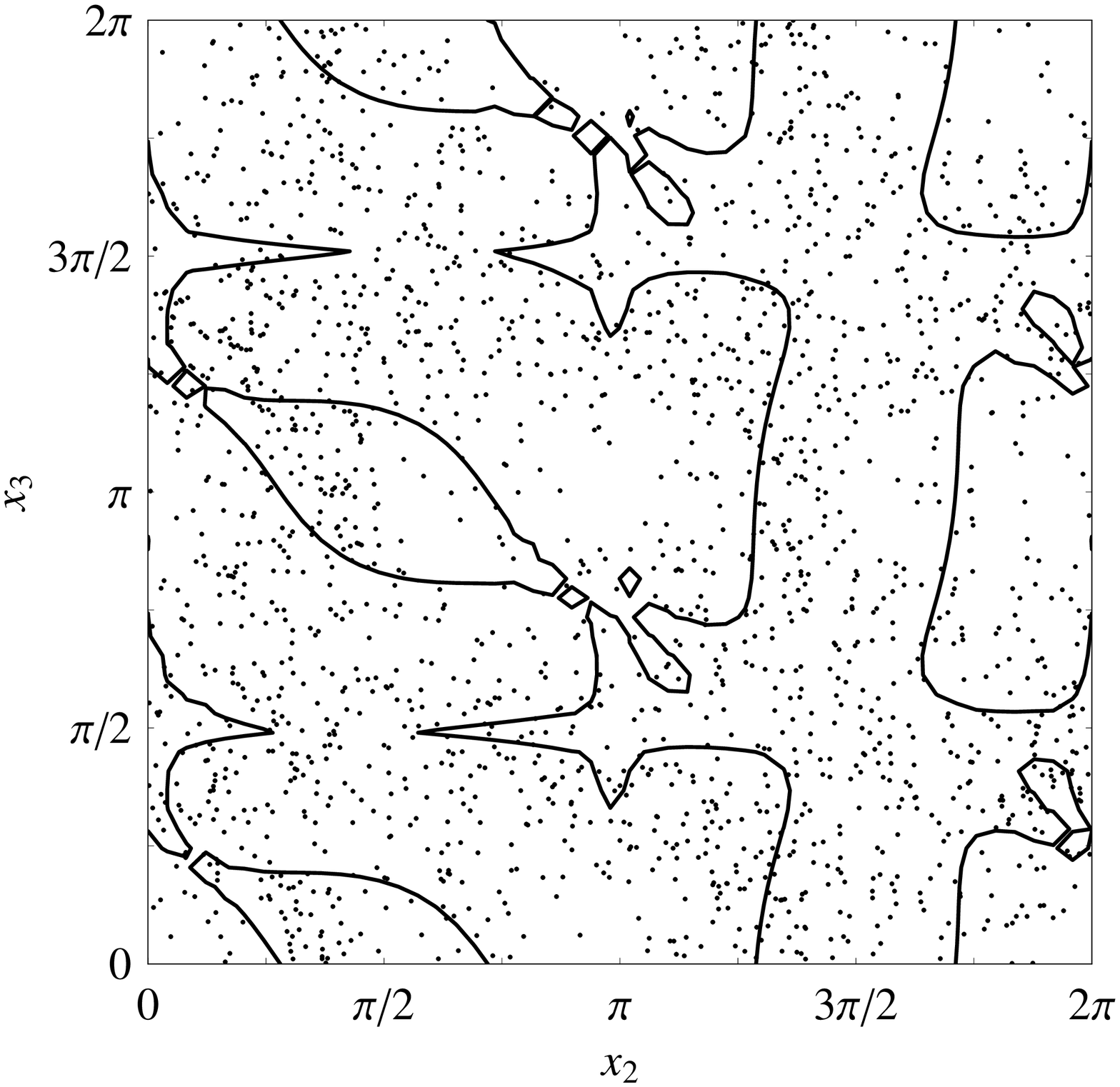}
 \caption{Regions in the plane ($x_2$, $x_3$) consistent with the generation of
       the baryon asymmetry via leptogenesis and the gravitino
       bound $M_1=10^{10,11,12}$~GeV, from left to right, respectively.
       The procedure yielding the scatter plots is
       described in
       the text.
       Also shown are the contours of constant
       $M_1(x_2,x_3)=10^{10,11,12}$~GeV with all
       other parameters fixed (best fit values of oscillation
       parameters, $m_1=0$, vanishing CP phases, \(x_1=0\) and $y_i=0.2$,
\(i=1,2,3\)).
}
     \label{fig:x3_x2}
\end{figure}

\begin{figure}[t]
\centering
\includegraphics[clip,width=0.7\textwidth]{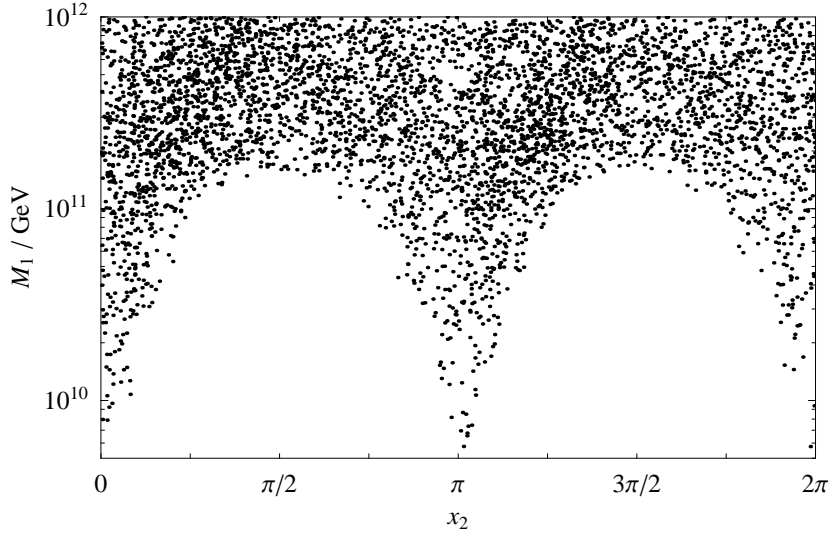}
 \caption{Region in the plane $(x_2, M_1)$ consistent with the
   generation of the baryon asymmetry via leptogenesis.
   The procedure yielding the scatter plot is described in the text.
}
     \label{fig:torboegen}
\end{figure}

The allowed ranges of \(x_2\) and \(x_3\) are shown in
Fig.~\ref{fig:x3_x2} for \(M_1=10^{10,11,12}\)~GeV (left to
right). All the other \(R\) matrix and light neutrino parameters
are scattered in their ranges as outlined above and as given in
Table~\ref{neutrino_parameters}, respectively. Each dot represents
a viable data point which respects all the above selection rules.
Fig.~\ref{fig:x3_x2} also shows contours of constant
\(M_1=10^{10,11,12}\)~GeV as illustration for which these
parameters are fixed at specific values. This behavior is also
nicely illustrated in Fig.~\ref{fig:torboegen}, which shows the
decrease of the allowed range of $x_2$ with decreasing $M_1$.
Above a value of $M_1 \approx 1.6\cdot 10^{11}$~GeV, $x_2$ can no
longer be constrained and the full parameter range is allowed. The
other $R$ matrix angle $x_3$ behaves analogously.

\section{Charged Lepton Flavor Violation}

In the previous sections the requirement of successful
leptogenesis has been used to constrain the lightest Majorana mass
$M_1$ and the parameters $y_i$, $x_{2,3}$ of the $R$ matrix. In
the SUSY version of the seesaw model further observables are
provided by the branching ratios of the radiative decays $l_j \to
l_i \gamma$. As has been shown in \cite{Deppisch:2002vz} in the
framework of degenerate Majorana neutrinos, these branching ratios
are particularly sensitive to the Majorana mass scale. In the
present framework of hierarchical Majorana neutrinos, a study of
$Br(l_j \to l_i \gamma)$ allows to obtain information about the
heaviest Majorana mass $M_3$.

In the mass insertion approximation, the branching ratio
\(Br\left(l_j\rightarrow l_i \gamma\right)\) is schematically
given by \cite{Hisano:1996cp,Casas:2001sr}
\beq
    Br\left(l_j\rightarrow l_i \gamma\right) \sim
        \frac{\alpha^3 \tan^2\beta}{\tilde{m}^8}
        \frac{m_{l_j}^{5}}{\Gamma_{j}}
        \left|\left(\delta m_{L}^2\right)_{ji}\right|^{2},
        \label{mass_insertion_formula}
\eeq
where \(\tilde{m}\) denotes the typical mass scale of the
sleptons in the loop. For our numerical calculations, we use the
complete one-loop results for $Br\left(l_j\rightarrow l_i
\gamma\right)$ \cite{Hisano:1996cp,RGE}.

Some general conclusions can be drawn from the dependence on the
five relevant, unknown seesaw parameters
(see (\ref{yprox}), (\ref{eq:rnrges}))
$M_3$, $x_{1,2}$ and $y_{1,2}$,
\beq
(\delta m_L^2)_{ij} \propto
\left(Y_{\nu}^{\dagger} L Y_\nu\right)_{ij} \propto M_3 |\cos(x_2
+ i y_2)|^2 \log\frac{M_X}{M_3}f_{ij}(x_1+i y_1), \label{yydag}
\eeq
where
\bea f_{12}&=&\left(\sqrt{\frac{m_2}{m_3}}\hat{s}_1
V_{12}^* + \hat{c}_1 V_{13}^*\right)^*
 \left(\sqrt{\frac{m_2}{m_3}}\hat{s}_1 V_{22}^* + \hat{c}_1 V_{23}^*\right),
\label{ffun1}\\
f_{23}&=&\left(\sqrt{\frac{m_2}{m_3}}\hat{s}_1 V_{22}^* + \hat{c}_1
V_{23}^*\right)^*
 \left(\sqrt{\frac{m_2}{m_3}}\hat{s}_1 V_{32}^* + \hat{c}_1 V_{33}^*\right).
\label{ffun2}
\eea

\begin{figure}[t]
\centering
\includegraphics[clip,width=0.48\textwidth]{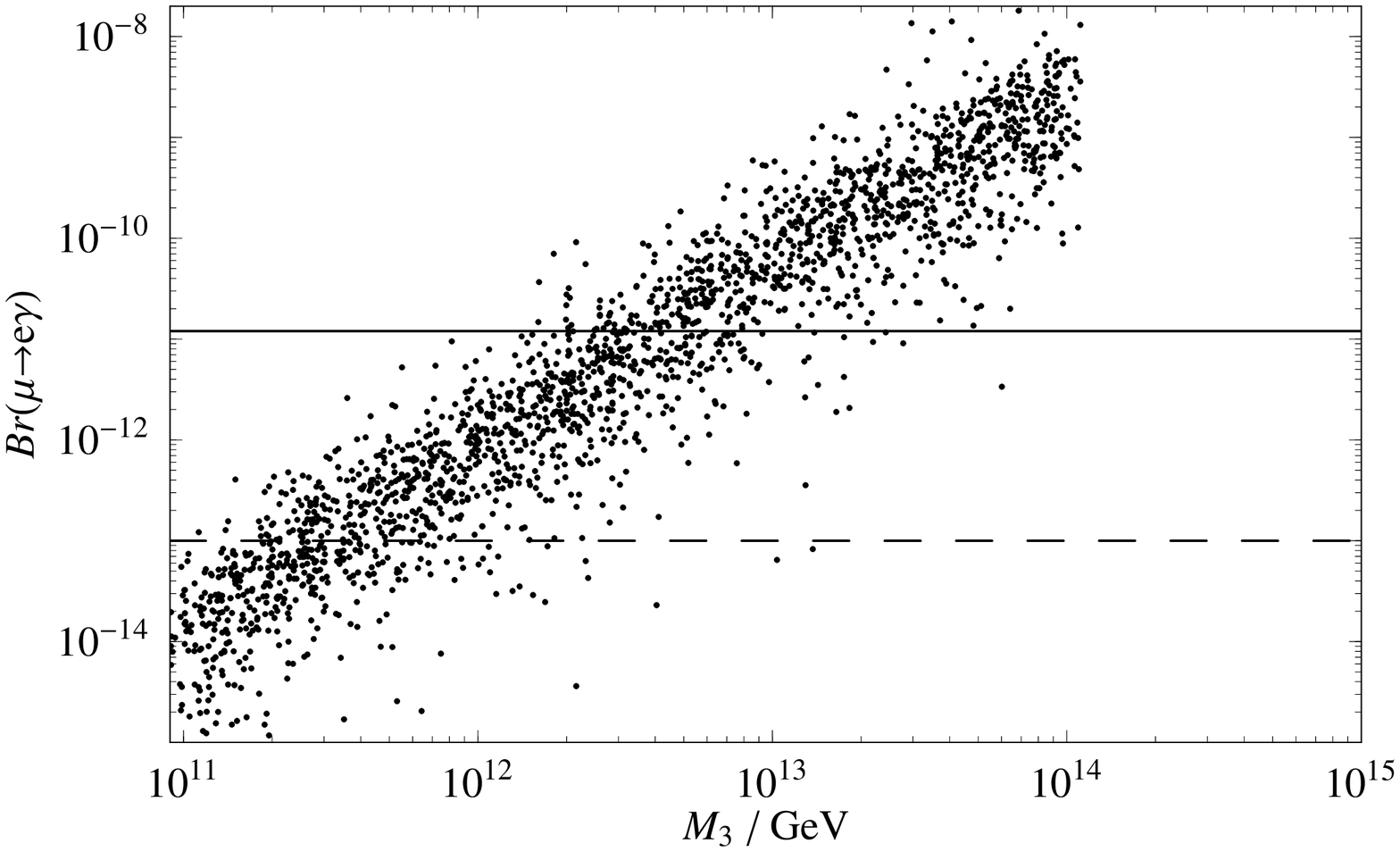}
\includegraphics[clip,width=0.48\textwidth]{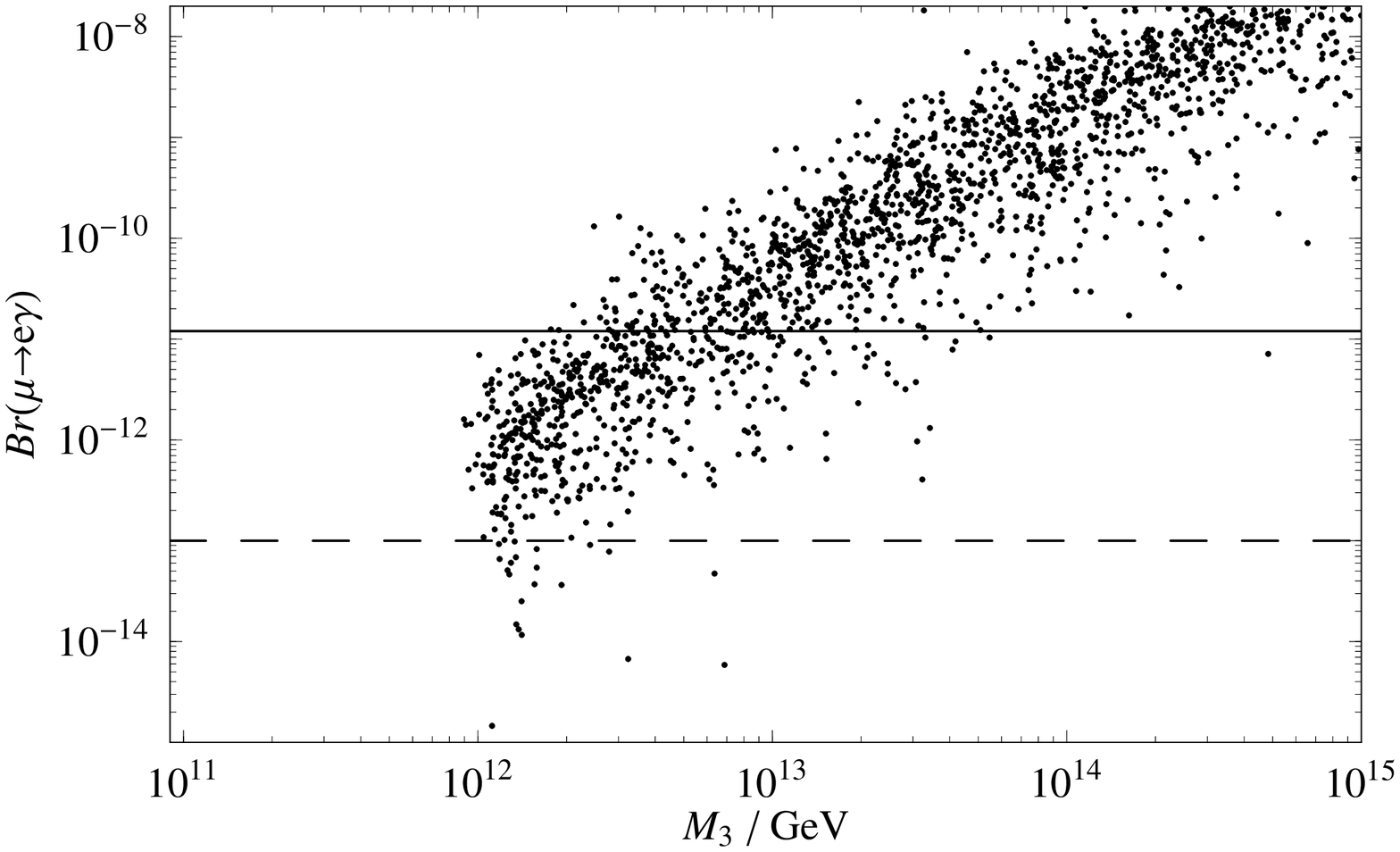}\\
\includegraphics[clip,width=0.48\textwidth]{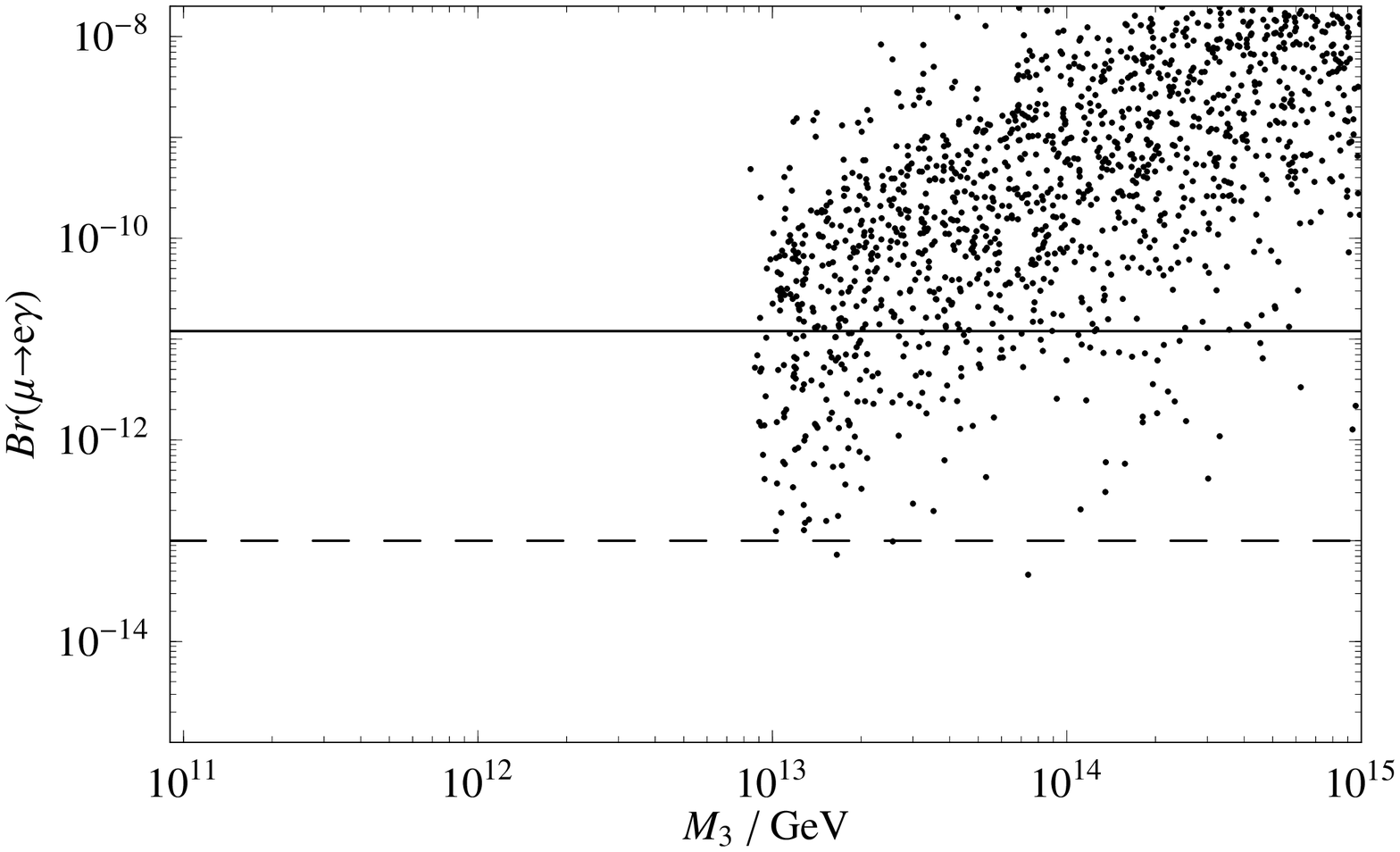}
     \caption{Branching ratio of $\mu \to e \gamma$ as a function of the
             heaviest Majorana mass $M_3$ in SUSY scenario SPS1a, for
            $M_1=10^{10,11,12}$~GeV, respectively. The solid
            (dashed) line indicates the present (expected future) experimental
            sensitivity. For a description of the scatter procedure
            see text.
}
     \label{fig:br12_m3}
\end{figure}

These expressions can be contrasted with the present experimental limits
\cite{megpres,tmg},
\bea
Br(\mu \to e \gamma)&<&1.2\cdot 10^{-11},\label{eqn:emu_present} \\
Br(\tau \to \mu \gamma)&<&6.8\cdot 10^{-8},\label{eqn:taumu_present}
\eea
as well as the expected sensitivity of the MEG experiment at
PSI \cite{meg2} and data from future B-factories \cite{tmg},
\bea
Br(\mu \to e \gamma)&\simeq& 10^{-13},\label{eqn:emu_future} \\
Br(\tau \to \mu \gamma)&\simeq& 10^{-8}.\label{eqn:taumu_future}
\eea
From (\ref{mass_insertion_formula},\ref{yydag})
it is immediately obvious that the branching ratios
increase with the square of the heaviest Majorana mass $M_3$.
This behavior is illustrated in Fig.~\ref{fig:br12_m3} for
\(M_1=10^{10,11,12}\)~GeV (left to right)
and provides a
possibility to determine $M_3$ from the combination of parameters
$M_3 |\cos\theta_2|^2 \simeq M_3$. The present bound on $Br(\mu \to e \gamma)$
constrains $M_3$ already to be smaller than some $10^{13}$~GeV.
The sensitivity of the future MEG experiment at PSI
would allow to bound
$M_3<{\cal O}(10^{12})$~GeV or determine the value of $M_3$ with an
accuracy of a factor of 10. Due to the hierarchy \(M_3/M_1\geq10\) used in our
scattering, the
minimum value for \(M_3\) rises with \(M_1\), pushing more and more points above
the present
limit on the branching ratio.
The widening of the scatter range for \(M_1<10^{12}\)~GeV (third plot) is
largely due
to the fact that \(x_2\) is no longer constrained around \(n\pi\), as can be
seen in the third plot of
Fig.~\ref{fig:x3_x2}. Hence, \(Br(\mu\to e\gamma)\propto|\cos\theta_2|^2\) can
be strongly suppressed.

\begin{figure}[t]
\centering
\includegraphics[clip,scale=0.50]{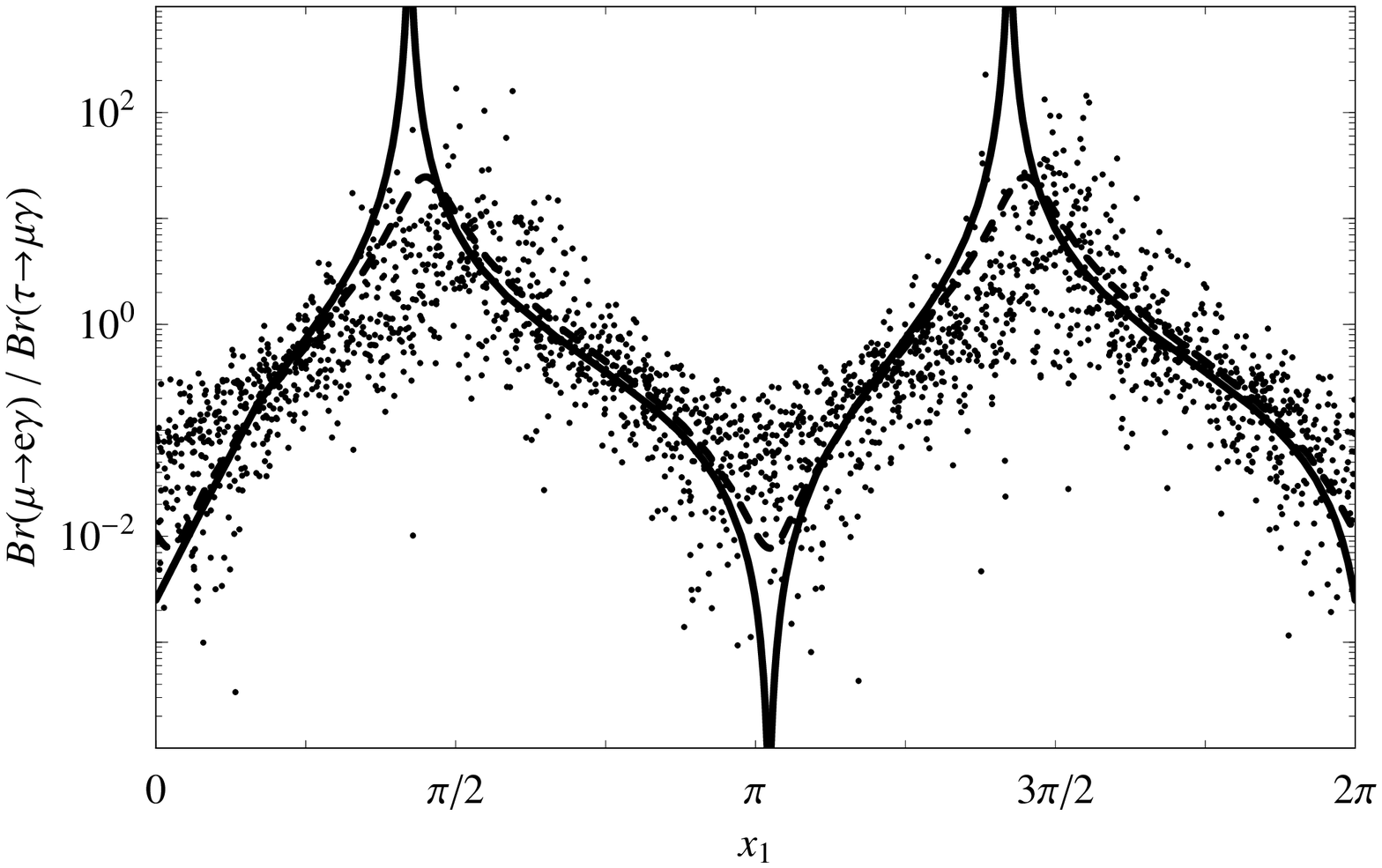}
     \caption{Ratio Br($\mu \to e \gamma$)/Br($\tau \to \mu \gamma$)
         as a function of $x_1$ in SUSY scenario SPS1a and for
         $M_1=10^{11}$~GeV.
         For a description of the scatter procedure
            see text. Superimposed are two curves for
       which all parameters other than $x_1$ are fixed (best fit oscillation
       parameters, $m_1=0$~eV, no CP phases and $y_{2,3}=0$).
       The solid (dashed)
       curve corresponds to $y_1=0.01$ ($y_1=0.1$).
}
     \label{fig:br12_br23}
\end{figure}

The dependence $Br(l_j \to l_i \gamma)$ on the initial and final
flavor is encoded in the functions $f_{ij}$, which show a
sensitive variation with the value of $x_1$. In particular,
$f_{12}$ and $f_{23}$ given in (\ref{ffun1},\ref{ffun2}) have
minima at small $y_1$ and different values of $x_1$, which imply a
strong variation of the ratio $Br(\mu \to e \gamma)/Br(\tau \to
\mu \gamma)$ with $x_1$. This ratio, which is rather independent
of the mSUGRA scenario used
(\ref{eq:rnrges},\ref{mass_insertion_formula}), is plotted in
Fig.~\ref{fig:br12_br23}. The dependence on $M_3 \cos^2\theta_2$
drops out and a measurement of both $Br(\mu \to e \gamma)$ and
$Br(\tau \to \mu \gamma)$ would allow to determine $x_1$ with an
accuracy of about ${\cal O}(1)$. It should be stressed, though,
that present (\ref{eqn:emu_present},\ref{eqn:taumu_present}) and
planned (\ref{eqn:emu_future},\ref{eqn:taumu_future}) experiments
have a very restricted sensitivity, $Br(\mu \to e \gamma)/Br(\tau
\to \mu \gamma)<{\cal O}(10^{-3})$. If future B-factories would
observe \(\tau\to\mu\gamma\), the angle \(x_1\) could also be
restricted to values \(x_1\approx n\pi\).

\section{Conclusions}

We have studied the minimal supersymmetric seesaw model by imposing
constraints from light neutrino data,  charged lepton flavor violation,
successful leptogenesis and perturbativity in the neutrino sector.
The parameters describing the
light neutrino sector have been fixed to their experimental best fit values
with an uncertainty anticipated from planned future experiments,
and a
very light neutrino spectrum, $m_1 \lsim m_2 \lsim m_3$ with
$m_1 < 0.03$~eV has been assumed.

The lightest right-handed neutrino mass
$M_1$ is fixed by the CP asymmetry $\epsilon_1$ imposed to generate the
observed value \(\eta_B=(6.3\pm0.3)\cdot 10^{-10}\) for the baryon asymmetry
and the
requirement of a small reheating temperature with
$M_1 < 10~T_R<{\cal O}(10^{11})$~GeV.
The largest Majorana mass $M_3$ is constrained from above
by the LFV process $\mu \to e \gamma$. Thus the right-handed neutrino
spectrum can be summarized as
\begin{equation}
{\cal O}(10^{11})~{\rm GeV} \sim M_1 < M_2 < M_3 \lsim
{\cal O}(10^{13})~{\rm GeV},
\end{equation}
in the given SUSY scenario SPS1a. Alternative SUSY scenarios with
sparticle masses in the range 100~GeV-1~TeV and low $\tan\beta$
are expected to yield
similar results.

In addition the successful scenario possesses an orthogonal $R$ matrix
encoding the mixing of the right-handed neutrinos, which is
parametrized by the
angles $\theta_i = x_i
+ i y_i$ with small but non-vanishing imaginary parts, being constrained as
\begin{eqnarray}
x_{2,3} &\simeq& n \pi,~~~~ n \in \N,\\
0 &<& y_{i} ~\lsim~ {\cal O}(1).
\end{eqnarray}
These bounds on $y_i$, $x_{2,3}$
result from the requirement of perturbative Yukawa couplings
and the condition of successful leptogenesis under the constraint
of a small $M_1$.
The remaining parameter $x_{1}$ can be determined from the ratio
$Br(\mu \to e \gamma)/Br(\tau \to \mu \gamma)$.

\section*{Note Added}
When this paper has been accomplished the preprint \cite{rode}
appeared, which comes to similar conclusions. One important
difference between the two studies is that the authors of
\cite{rode} made further theoretical assumptions leading to more
definite but less general results.

\section*{Acknowledgements}
We thank W. Buchm\"uller and M. Pl\"umacher for many useful
comments and discussions. This work was supported by the
Bundesministerium f\"ur Bildung und Forschung (BMBF, Bonn,
Germany) under the contract number 05HT1WWA2 and by US DOE under
the grant DE-FG03-91ER40833.

\end{document}